\documentclass[10pt,letterpaper]{article}
\usepackage[top=0.85in,left=2.75in,footskip=0.75in]{geometry}

\usepackage{amsmath,amssymb}

\usepackage{changepage}

\usepackage[utf8x]{inputenc}

\usepackage{textcomp,marvosym}

\usepackage{cite}

\usepackage{nameref,hyperref}

\usepackage{microtype}
\DisableLigatures[f]{encoding = *, family = * }

\usepackage[table]{xcolor}

\usepackage{array}

\newcolumntype{+}{!{\vrule width 2pt}}

\newlength\savedwidth

\raggedright
\usepackage[aboveskip=1pt,labelfont=bf,labelsep=period,justification=raggedright,singlelinecheck=off]{caption}

\makeatletter
\renewcommand{\@biblabel}[1]{\quad#1.}
\makeatother

\usepackage{lastpage,fancyhdr,graphicx}
\usepackage{epstopdf}
\fancyheadoffset[L]{2.25in}
\fancyfootoffset[L]{2.25in}
\usepackage{graphicx}
\usepackage{subfigure}
\usepackage{epstopdf}
\usepackage{times}
\usepackage{bm}
\usepackage{color}

\graphicspath{{figs/}}

\usepackage{float,amssymb}

\begin{document}

\vspace*{0.2in}

\begin{flushleft}
{\Large
\textbf\newline{Localization, epidemic transitions, and unpredictability of multistrain epidemics with an underlying genotype network} 
}
\newline
\\
Blake J. M. Williams\textsuperscript{1},
Guillaume St-Onge\textsuperscript{2,3},
Laurent H\'ebert-Dufresne\textsuperscript{1,2,4,*}
\\
\bigskip
\textbf{1} Vermont Complex Systems Center, University of Vermont, Burlington, Vermont, United States of America
\\
\textbf{2} D\'epartement de physique, de g\'enie physique et d'optique, Universit\'e Laval, Qu\'ebec, Canada
\\
\textbf{3} Centre interdisciplinaire en mod\'elisation math\'ematique, Universit\'e Laval, Qu\'ebec, Canada
\\
\textbf{4} Department of Computer Science, University of Vermont, Burlington, Vermont, United States of America
\\
\bigskip

* laurent.hebert-dufresne@uvm.edu

\end{flushleft}


\section*{Abstract}
Mathematical disease modelling has long operated under the assumption that any one infectious disease is caused by one transmissible pathogen spreading among a population. This paradigm has been useful in simplifying the biological reality of epidemics and has allowed the modelling community to focus on the complexity of other factors such as population structure and interventions. However, there is an increasing amount of evidence that the strain diversity of pathogens, and their interplay with the host immune system, can play a large role in shaping the dynamics of epidemics. Here, we introduce a disease model with an underlying genotype network to account for two important mechanisms. One, the disease can mutate along network pathways as it spreads in a host population. Two, the genotype network allows us to define a genetic distance across strains and therefore to model the transcendence of immunity often observed in real world pathogens. We study the emergence of epidemics in this model, through its epidemic phase transitions, and highlight the role of the genotype network in driving cyclicity of diseases, large scale fluctuations, sequential epidemic transitions, as well as localization around specific strains of the associated pathogen. More generally, our model illustrates the richness of behaviours that are possible even in well-mixed host populations once we consider strain diversity and go beyond the ``one disease equals one pathogen'' paradigm.

\section*{Author summary}
Epidemics rarely involve a single unique pathogen but are often modelled as such. Rather, most pathogens circulate under a family of strains which can interact differently with the host immune system and undergo further mutations. Here we extend a classic epidemiological model to consider the genetic structure connecting these strains --- i.e., the genotype network mapping possible mutation pathways --- and investigate the dynamics and emergence of epidemics beyond the ``one disease equals one pathogen'' paradigm. This simple model allows us to consider the impacts of (i) mutation, (ii) cross-immunity between strains, (iii) competition between strains, and (iv) the structure of the genotype network. We find that, altogether, these features do not affect the classic epidemic threshold but localize outbreaks around key strains and yield a second immune invasion threshold below which the epidemics follow almost cyclical and chaos-like dynamics. Our results illustrate how little biological realism is needed to introduce key features of real epidemics in even the simplest disease models: Epidemic cycles, unpredictability, and heterogeneous strain prevalence.

\section{Introduction}

Viral species are known to often undergo rapid evolution. Since the early 20th century, influenza viruses have been described as having marked variability and unpredictable behaviour \cite{Meiklejohn:1949}. Subsequent RNA virus studies of the 20th and 21st century have focused on, among others, the \textit{Zaire ebolavirus}, strains of the SARS-CoV species, and HIV-1, all possessing high mutation rates \cite{Sanjuan:2010}. These frequent mutations contribute to the antigenic evolution of these viruses, allowing them to evade recognition by the human immune system \cite{deJong:2007}.

Despite the long-standing knowledge of subtypes and strains within viral species, mathematical disease modelling has continued to model viral diseases with one underlying pathogen. Notably, influenza violates the ``one disease, one pathogen'' paradigm: numerous types, subtypes, and strains of influenza viruses challenge the human immune system, driving vaccine effectiveness below 50$\%$ in most recent years \cite{flannery:2015,flannery:2016,flannery:2017,Doyle:2018}. Models which fail to account for antigenic variation of a pathogen may lead to biased characterizations of epidemic emergence and progression.

Modelling multi-strain pathogens with consideration for antigenic properties requires the inclusion of cross-protective effects, in which the immunity acquired towards one strain offers partial protection towards another strain based on their antigenic similarity. Cross-protection is seen in numerous viral species \cite{Gilchuk:2016, Hensley:2010,Epstein:2010}. In general, more similar strains will have greater cross-protective effects, as with seasonal influenza \cite{Peeters:2017}. However, cross-protective immunity is not necessarily a monotonically decreasing function of antigenic distance. Antibody-dependent enhancement has been observed in dengue viruses, in which a past infection may in fact increase the risk of severe infection \cite{Halstead:2003,Katzelnick:2017}. Regardless, approximations of cross-protection may be made through antigenic distance or genetic distance. This relationship may be determined by a function of genetic distance to approximate the unique antigenic distances between all strains. 

Several models have been proposed in the growing sub-discipline of multistrain disease modelling \cite{Kucharski:2015}. These models balance biological assumptions with computational tractability through reduction via symmetry (e.g. antigenic neighbourhoods \cite{Ferguson:2002}), age structure \cite{Kucharski:2012}, and deciding to capture either infection history or immune status \cite{Kucharski:2015}, among other modelling choices. Cross-protective immunity has been explored in two-strain models \cite{Kamo:2002}, multi-strain models with a restricted number of antigenic loci and alleles \cite{Minayev:2009}, and temporary cross-protective immunity in dengue models \cite{feng1997competitive} capable of producing cyclical and chaos-like infection progression.  However, the effects of an underlying genotype network structure --- governing viable mutation pathways and genetic distances between strains --- have not been thoroughly explored with multistrain models. Genotype networks consist of nodes that represent strains, with edges connecting strains that differ by one nucleotide or amino acid in some antigenic region of a gene or protein \cite{Wagner:2014}. Genotype networks are a complementary structure to phylogenetic trees, and are a useful way of representing genetic distance necessary for cross-immunity in multi-strain models.

Moreover, the genotype network gives us a proxy through which we can specify potential mutation pathways between strains. Mechanisms for pathogen mutation have previously been included in mathematical models\cite{girvan2002simple,uekermann2012spreading}, often to consider the emergence of antiviral resistance \cite{lipsitch2007antiviral,hebert2013pathogen,althouse2013timing,patterson2013optimizing}. Particularly, these models predict the emergence of sequential epidemic transitions --- with a first epidemic threshold defining the emergence of macroscopic disease spread and a second marking the emergence of treatment resistant strain \cite{hebert2013pathogen}. However, such models are often limited to only two pathogen strains as they require specification of the fitness cost associated with resistance. We therefore aim to introduce a more general model, allowing large number of strains to mutate along specific network pathways. While this general model could consider a complex fitness landscape over this genotype network, we focus on the case of neutral evolution and show how the previous results discussed here can all co-exist within a single, fairly simple model.

We introduce a multistrain Susceptible-Infectious-Recovered-Susceptible (multistrain SIRS) epidemic model with an underlying genotype network, allowing the disease to evolve along plausible mutation pathways as it spreads in a well-mixed population. We then investigate the effects of genotype network structure on the emergence of an endemic state and on the fitness distribution of strains across the genotype network. Altogether, our results challenge the typical phenomenology of epidemic models.
We observe localization of infections in the genotype space and identify two epidemic transitions. The first corresponds to the emergence of an endemic state where new infections mostly target the susceptible portion of the population, and the second marks the point where recovered individuals significantly contribute to new infections. Between these thresholds, we find chaos-like behaviour which can be maintained for arbitrarily long times, yielding time series with epidemic cycles featuring large, unpredictable fluctuations.

\section{Methods}

We study the spread of infectious disease within a well-mixed population for a defined genotype network of the chosen pathogen. Our model is as follows.

The underlying epidemiological dynamics correspond to a simple SIRS model, but where we add a genotype network defined as a set of potential mutations, meaning an infection of strain $i \in [1,N]$ can mutate along the network to a neighbouring strain $j\in \mathcal{N}_i$, where $\mathcal{N}_i$ specifies the set of first network neighbours of strain $i$. Biologically, this network is defined such that neighbouring strains $i$ and $j$ differ by one unit of genetic distance. 

\begin{figure}[]
\centering
\includegraphics[width=\linewidth]{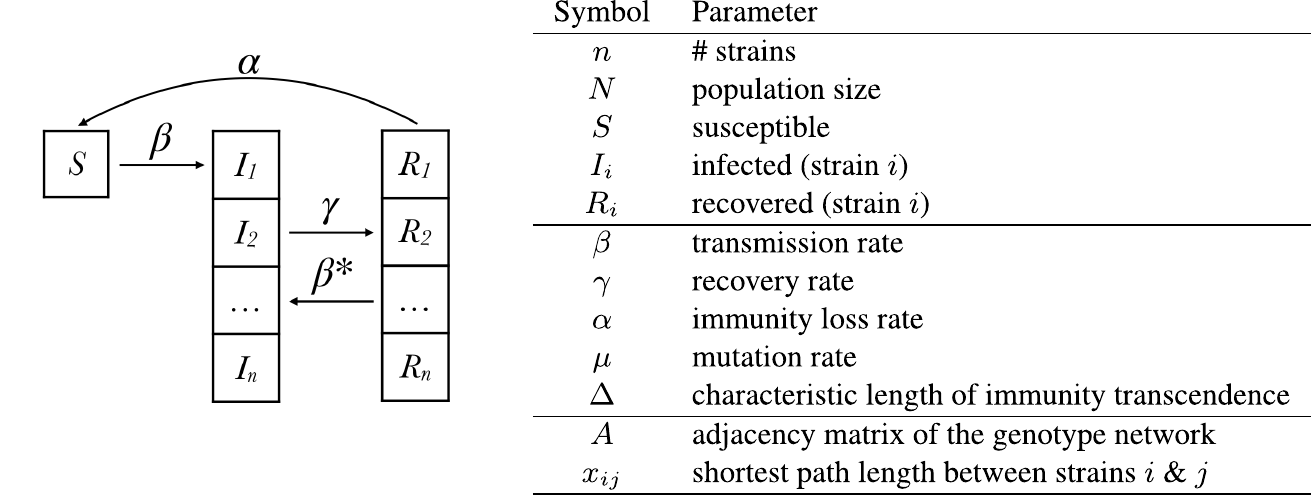}
\caption{Compartmental model of a multistrain epidemic with an underlying genotype network \textbf{(left)} and table of parameters \textbf{(right)}.}
\label{fig:model}
\end{figure}

The strains spread within a well-mixed host population. Host individuals are defined as susceptible ($S$) if they possess no immunity to any strain of a disease, see Fig \ref{fig:model}. Susceptible individuals progress to infectious state $I_i$ at transmission rate $\beta$ for every contact with individuals infectious with strain $i$, occurring at rate $\beta I_i$ for every susceptible individual. Note that this basic transmission rate is held constant for all strains, as we focus on neutral evolution as a first approximation. 

Individuals in $I_i$ can either: (i) recover at rate $\gamma$ to state $R_i$ and acquire direct immunity for strain $i$ and partial immunity to strain $j\neq i$; or (ii) become infected with strain $I_j$ via mutation at a rate $\mu$ for all strains $j$ in $\mathcal{N}_i$. Individuals in $R_i$ will either: (i) lose immunity and progress to $S$ at rate $\alpha$, or (ii) become infected with strain $j\neq i$ to which they only possessed partial immunity and progress to $I_j$ at a reduced rate $\beta^*$, where $\beta^*$ is an exponentially decaying function of genetic distance between strains $i,j$.  Specifically:
\begin{equation}
\beta^* \propto 1-e^{-x_{ij}/\Delta}
\end{equation}
where $x_{ij}$ is the genetic distance between strain $i,j$ (approximated by shortest path of length $x_{ij}=x_{ji}$ between strains $i,j$ in the genotype network) and $\Delta$ is the characteristic length of immunity transcendence ($0<\Delta<\infty$). This model makes the simplifying assumption that an individual only possesses immunity to the most recent strain of infection, existing in one immune state at a time. 

Altogether, the dynamics of our model can be followed by the following set of ordinary differential equations (ODEs),
\begin{align}
    \frac{dS}{dt} &= -\beta \sum_{i=1}^{N} \frac{SI_i}{N} + \alpha \sum_{i=1}^{N}R_i \label{eq1}\\
    \frac{dI_i}{dt} &= \beta \frac{SI_i}{N} - \gamma I_i + \mu \sum_{j=1}^N A_{i,j}(I_j-I_i) + \beta \sum_{j=1}^N \left(1-e^{-x_{ij}/\Delta}\right) \frac{I_iR_j}{N} \label{eq2}\\
    \frac{dR_i}{dt} &= \gamma I_i -\alpha R_i - \beta \sum_{j=1}^N \left(1-e^{-x_{ij}/\Delta}\right) \frac{I_jR_i}{N} \label{eq3}
\end{align}
where $A_{ij}$ is an element of the adjacency matrix of the genotype network, equal to 1 if there is mutation pathway between $i$ and $j$ and 0 otherwise. The total proportion of individuals infected can be obtained by summing over all strains, $I(t) = \sum_i I_i(t)$, and we also focus on its asymptotic value $I^* = \lim _{t\rightarrow \infty}\sum_i I_i(t)$.

We therefore have 5 important epidemiological parameters: transmission rate $\beta$, recovery rate $\gamma$, rate of waning immunity $\alpha$, mutation rate $\mu$ and immunity transcendence $\Delta$. Unless mentioned otherwise, we fix the recovery rate $\gamma = 1$ such that other rates are defined in units of infectious period (and $\Delta$ in units of genetic distance). The other parameters then allow us to investigate different regimes of interest.

As $\Delta \rightarrow 0$, immunity becomes strain specific with no cross-protective effects. As $\Delta \rightarrow \infty$, immunity becomes broad-reaching to the point of universal protection across all strains. With $\beta^*$ as a function of distance, we are able to reduce model complexity by avoiding specification of $\beta^*_{ij}$ among all strains, whose values may not be known in real-world applications. Instead, we rely on the inverse relationship between antigenic distance and cross-protection that has been observed in influenza viruses \cite{Peeters:2017}. Note that this relationship may not be monotonically decreasing for all pathogens, in which case $\beta^*$ may be defined by a function of genetic distance unique to the pathogens.

Our most important assumption is perhaps that only the most recent infection is relevant for cross-immunity effects. Indeed, $I_i$ and $R_i$ specify the pathogen involved only in the most recent infection for every individual. The alternative would have been to model an infectious state $I_{i,j,...}$ for all unique infection histories, of complexity $\mathcal{O}(2^n)$ if order does not matter and complexity $\mathcal{O}(n!)$ if it does. While a big assumption, focusing on the most recent infection reduces the complexity to $\mathcal{O}(n)$. Computational feasibility would be largely restricted otherwise, which would limit the analysis of the effects of genotype network structure \cite{Gomes:2002,Kucharski:2015}. The infection history approximation enables genotype networks to be large enough to contain complex structure, necessary to investigate the role of genotype networks in epidemic progression.

\section{Results}

We focus our attention on the consequences of the genotype network underlying the spread of the disease. In order to gain as much insights as possible on how it affect prevalence of a disease, we keep the network itself simple using well-known graph toy models composed of lattices, chains and stars.

\subsection{Localization in genotype space}

We first ask which strains can be expected to have an advantage, not because of their own fitness or of our epidemiological parameters (as they all share the same $\beta$, $\gamma$ and $\alpha$), but because of their position in the genotype network. We use three simple network structures --- a star, a square lattice, and a chain, all containing 25 strains --- and run our model to produce a large outbreak with $\beta = 25$ much greater than the expected SIRS epidemics threshold of $\beta_c = 1$. Accordingly, we set the evolutionary dynamics to be much slower than that of epidemic spread with $\mu = 10^{-3}$. We then let the system reach its endemic steady state, where the derivatives in Eqs.~(\ref{eq1}-\ref{eq3}) essentially go to zero such that the system is at equilibrium.

\begin{figure}[t!]
    \centering
    \includegraphics[width=\linewidth]{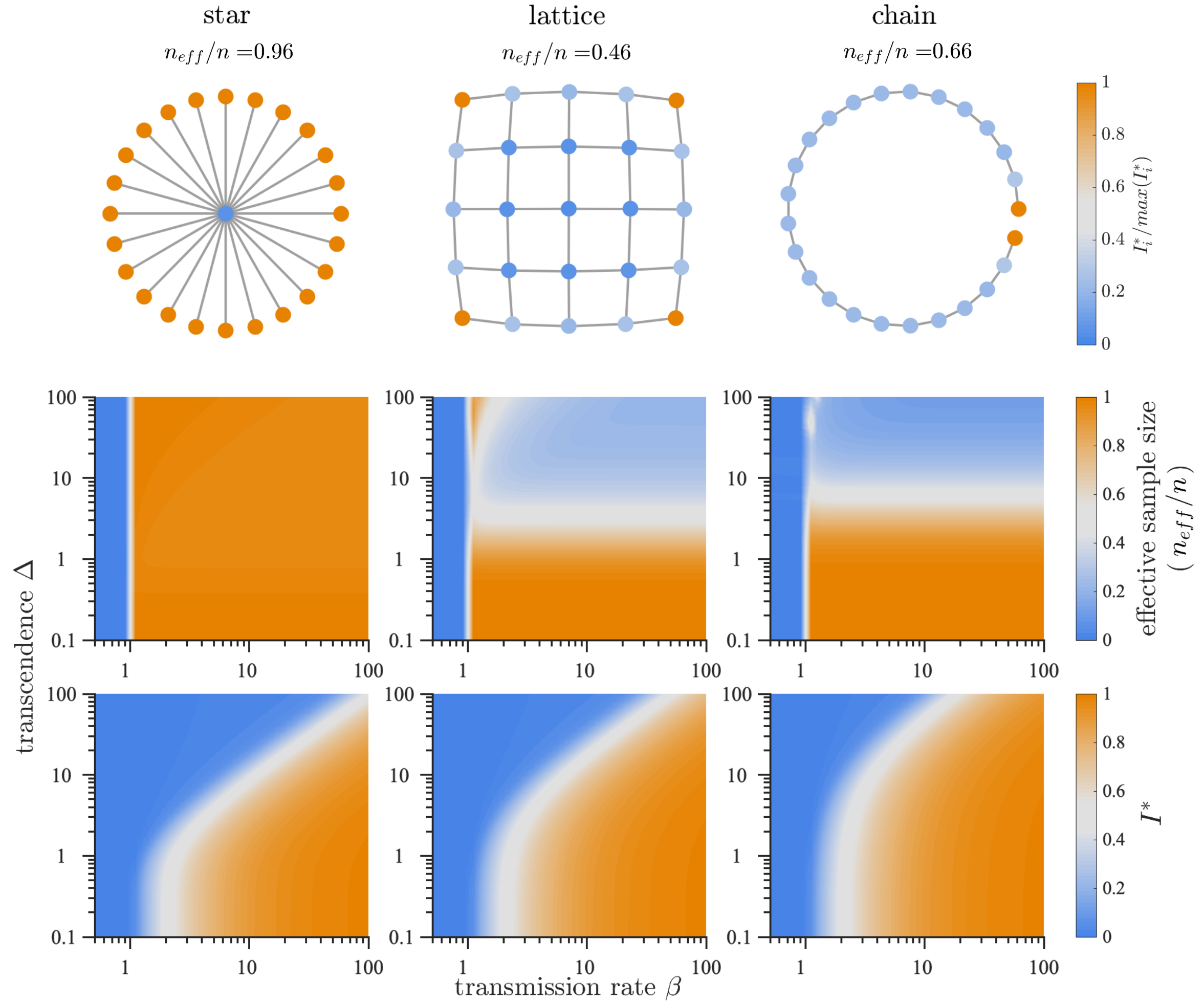}
    \caption{Infection localization and characteristics of endemic infection state. \textbf{(Top)} Localization within networks shown by endemic infection counts $I_i^*$ normalized by max($I_i^*$) for a given network. We use mutation rate $\mu=10^{-3}$,  transmission rate $\beta = 25$, waning immunity rate $\alpha=1/50$, and transcending immunity $\Delta=4$. \textbf{(Middle)} Infection localization regimes are revealed where normalized effective sample size is low (lattice and chain), occurring when few strains account for the majority of infections. \textbf{(Bottom)} endemic infections depend on not only transmission rate $\beta$, but also the breadth of cross-protective effects determined by transcendence rate $\Delta$. Fixed parameters are $n=25$, $\mu=10^{-3}$, $\alpha=1/50$.} 
    \label{fig:localization_heat}
\end{figure}

We observe a localization of infections by strain within genotype networks as shown in Fig \ref{fig:localization_heat}, top row. Stationary or endemic infection counts differ from strain to strain. This holds true even with the assumption of neutral evolution, based solely on their position in the network and the resulting cross-protective immune effects. Epidemics can therefore be localized around a minority of strains, as is clear in the lattice and chain networks.

We quantify this localization phenomena with Kish's effective sample size \cite{Kish:1965}, referred to here as effective participation ratio $n_{eff}^*=n_{eff}/n=\left(\sum I_i \right)^2 / \left(n\sum I_i^2\right)$. As $n_{eff}^* \rightarrow 1$, all strains contribute an equal number of infections. As $n_{eff}^* \rightarrow n^{-1}$, only one strain contributes infections. In Fig \ref{fig:localization_heat}, top row, we observe lower $n_{eff}^*$ in the lattice and chain, indicating greater localization. A small number of strains are able to escape strong cross-protective immunity in the corners of the lattice and at the ends of the chain, while such heterogeneity is not seen in the star and ring networks.

As network structure determines infection localization, so does the transcendence of immunity. In Fig \ref{fig:localization_heat}, middle row, we see $n_{eff}^*$ as a function of $\beta$ and immunity transcendence $\Delta$, revealing regimes of strong localization in the lattice and chain networks where $n_{eff}^*$ remains small. High values of $\Delta>10$, indicating far-reaching cross-protection, are associated with localization in these two networks. The structure of the star and loop networks allow them to escape localization effects influenced by large $\Delta$.

Stationary infection counts $I^*$ are also influenced by immunity transcendence $\Delta$. In Fig \ref{fig:localization_heat}, bottom row, we see reductions in $I^*$ as $\Delta$ increases. As cross-protective effects increase, a higher $\beta$ becomes necessary to maintain infections. Again we see the importance of network structure, with different values of $\Delta$ required between networks to affect $I^*$.

\subsection{Sequential phase transitions}

We then look at the behaviour of the endemic state as we vary the basic transmission rate $\beta$. We know from classic SIRS model that there should be an epidemic transition at $\beta_c=1$, marking a transition between a disease-free phase where the disease is too weak to establish itself in the population if $\beta<1$, and an endemic phase for larger values. Yet, one interesting result of Fig \ref{fig:localization_heat}, bottom row is that the epidemic threshold now seem to increase with transcending immunity $\Delta$. This is somewhat surprising given that $\Delta$ does not matter for any one strain, which should still be able to survive on its own following SIRS dynamics once $\beta > \beta_c = 1$.

\begin{figure}[t!]
    \centering
    \includegraphics[width=0.99\linewidth]{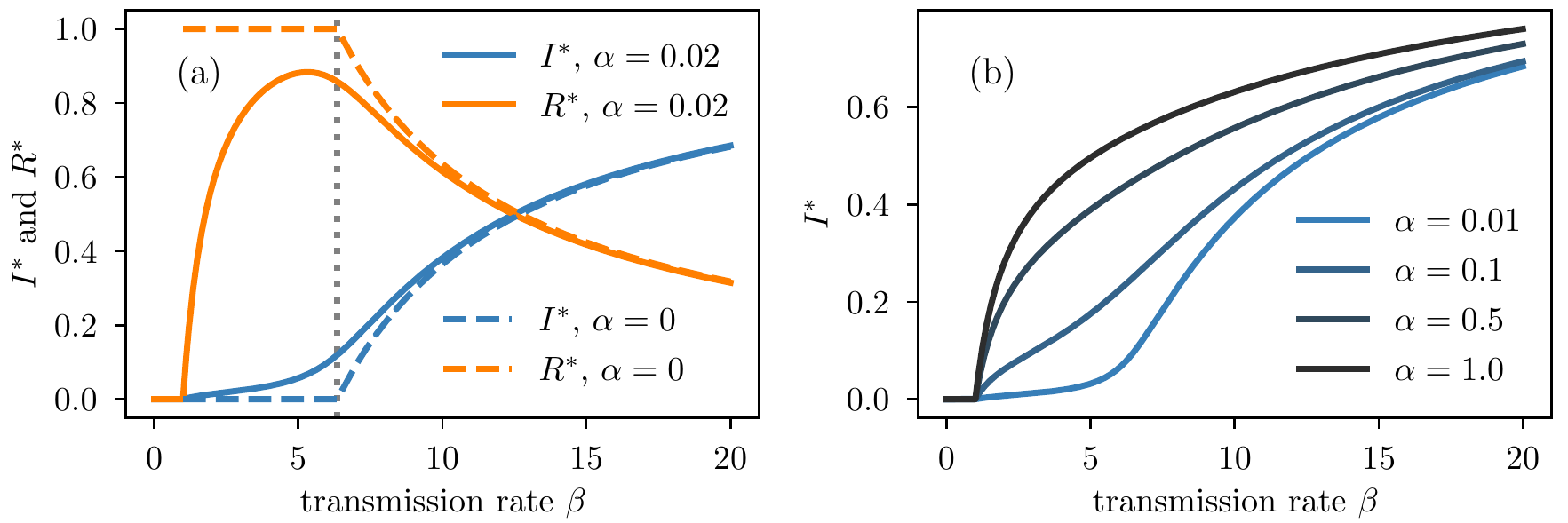}
    \caption{Bifurcation diagram for the model with varying levels of waning immunity on a star genotype network with 10 strains. We fix the recovery rate to $\gamma = 1$, the mutation rate $\mu = 1/100$, the transcending immunity $\Delta = 10$ and we vary the transmission rate $\beta$. \textbf{(a)} We set the waning immunity rate $\alpha \in \left \{0, 0.02\right\}$ to illustrate the origin of the immune invasion threshold (vertical dotted line) obtained with Eqs.~\eqref{eq:stationary_I_no_waning} and \eqref{eq:immune_invasion_threshold}. \textbf{(b)} For large enough values of waning immunity rate, the immune invasion threshold disappears because recovered nodes quickly become susceptible again. }
    \label{fig:bifurcation_star_10}
\end{figure}

In Fig \ref{fig:bifurcation_star_10}, we take a deeper look at the phase diagram under varying transmission rate and observe a \textit{second} epidemic transition. More precisely, if $\alpha$ is not too large, $I^*$ is no longer a concave function of the transmission rate; it emerges as expected at $\beta_c = 1$ but has a new inflection point at a much higher $\beta$ value. This means that only modest increases in $I^*$ are seen when $\beta$ is just above to the epidemic threshold $\beta_c=1$, in contrast to standard SIS-like models in which this regime experiences the most rapid rate of change in $I^*$ as a function of $\beta$ \cite{Pastor:2015}. 

We conjecture that this second phase transition is governed by what we call the \textit{immune invasion threshold}, corresponding to the point at which infected nodes starts to infect recovered nodes (of other strains) effectively. To see this, in Fig \ref{fig:bifurcation_star_10} (a), we compare the bifurcation diagrams of two models: with and without waning immunity. In the latter case, in the stationary state, a node infected with strain $i$ can only infect recovered nodes of strains $j \neq i$ (since $S^* = 0$). The immune invasion threshold $\beta_\mathrm{I}$ can thus be estimated from $\beta_\mathrm{c}$ if $\alpha \mapsto 0$. Surprisingly, even though $\beta_\mathrm{c} = 1$ whenever $\alpha > 0$, it is no longer the case when $\alpha = 0$.

Previous work has demonstrated a similar phenomenon in which $I^*$ as a function of $\beta$ is dependent upon the level of transcendence of immunity \cite{Andreasen:Sasaki, Omori}. This indicates the importance of transcending immunity in multi-strain modelling. However, our model differs by showing the development of a second transition that is absent from these models, having potential consequences regarding the robustness of $I^*$ with change in $\beta$. The second transition reveals that little changes may occur in $I^*$ below the second transition, while $I^*$ may change rapidly above this transition in our model. Furthermore, the structure of the genotype network itself is critical to the nature of this second transition. This difference with previous work could be consequential in evaluating the utility of public health interventions aimed at reducing transmission rate.

To derive the immune invasion threshold, let us rewrite the stationary state quantities $I^*,R^*$ when $\alpha = 0$. We have
\begin{align*}
    0 &= -\gamma I_i^* + \mu \sum_j A_{ij} I_j^*-\mu k_i + \beta I_i^* \sum_j T_{ij} R_j^* \;, \\
    0 &= \gamma I_i^* - \beta R_i^* \sum_j T_{ij} I_j^* \;,
\end{align*}
where $T_{ij} \equiv (1 - e^{-x_{ij}/\Delta})$. Isolating $R_i^*$ in the second equation and reinjecting the solution in the first one, we obtain a self-consistent equation for the $\left \{I_i^* \right \}$,
\begin{align}\label{eq:stationary_I_no_waning}
    I_i^* &= \frac{\sum_{j}A_{ij} I_j^*}{\frac{\gamma}{\mu}\left( 1 -\sum_j T_{ij} \frac{I_j^*}{\sum_k T_{jk} I_k^*}\right) + k_i} \;.
\end{align}

Interestingly, the $\left \{I_i^* \right \}$ do not depend upon $\beta$. However, we know that such solution is possible only if $I_i^* > 0 \; \forall i$, and this break down at $\beta_\mathrm{I}$ when $R^* = \sum_i R_i^* \to 1$. Therefore, the immune invasion threshold $\beta_\mathrm{I}$ has the following explicit expression
\begin{align}\label{eq:immune_invasion_threshold}
    \beta_\mathrm{I} = \gamma \sum_i \frac{I_i^*}{\sum_j T_{ij} I_j^*} \;,
\end{align}
where the $\left \{I_i^* \right \}$ are evaluated from Eq.~\eqref{eq:stationary_I_no_waning}.

We observe a direct relationship between $\Delta$ and $\beta_\mathrm{I}$. 
Namely, when $\Delta \to \infty$, $T_{ij} \to 0$ for all $i,j$, hence $\beta_\mathrm{I} \to \infty$, as seen from Eq.~\eqref{eq:immune_invasion_threshold}. When $\Delta \to 0$, $T_{ij} \to 1$ for all $i \neq j$, and $T_{ii} = 0$, hence
\begin{align*}
\beta_\mathrm{I} \to \gamma \sum_i \frac{I_i^*}{\sum_{j \neq i} I_j*} \;.
\end{align*}
For large networks, $\beta_\mathrm{I} \approx \gamma \equiv 1$ in the limit $\Delta \to 0$.
Therefore, we conclude that increasing $\Delta$ increases the immune invasion threshold, which makes sense based on intuition alone.

\begin{figure}[]
    \centering
    \includegraphics[width=\linewidth]{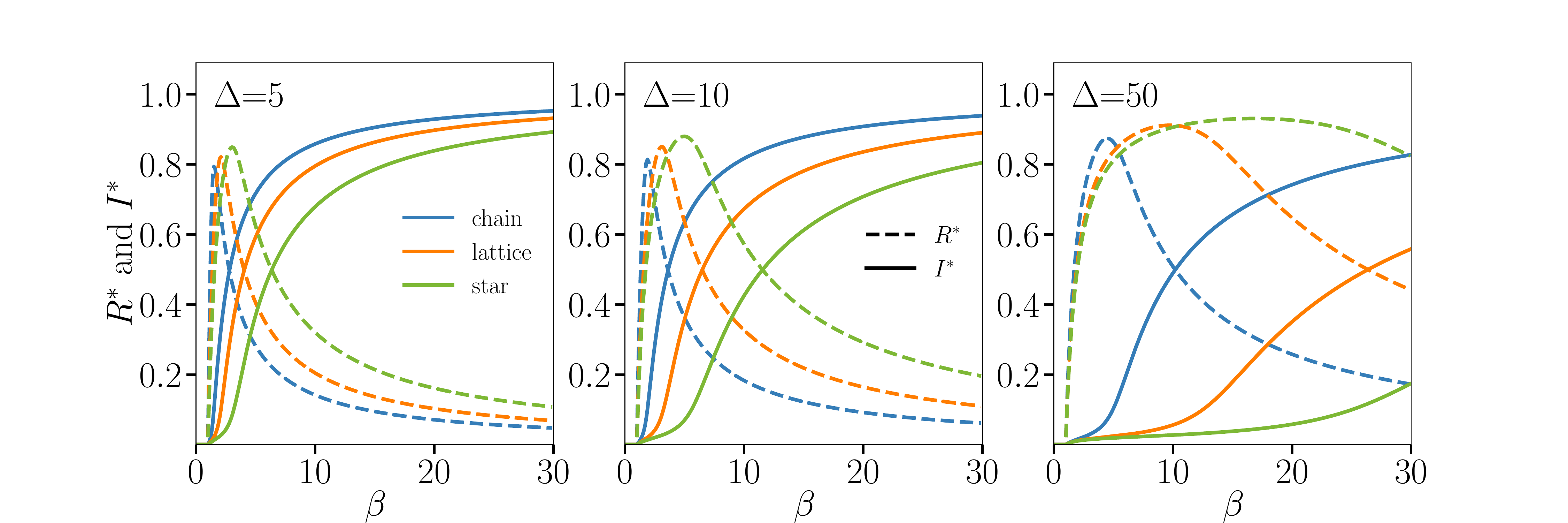}
    \caption{Integration of the ODEs on three toy genotype networks --- the chain, square lattice and star --- all with 25 strains. We fix the recovery rate $\gamma=1$, the mutation rate $\mu=1/100$, the waning immunity rate $\alpha =1/50$ and vary the transmission rate $\beta$ under three values of transcending immunity: $\Delta = 5$ \textbf{(Left)}, $\Delta = 10$ \textbf{(Center)}, and $\Delta = 50$ \textbf{(Right)}. Close to the inflection point of every $I^*$ curve (shown in solid lines) we find a maximum in $R^*$ (shown in dashed lines). This point therefore marks a second activation threshold, one where the transmission rate is high enough to counteract transcending immunity and spread the outbreak using the pool of recovered individuals.}
    \label{fig:2nd_threshold}
\end{figure}

The relationship between the immune invasion threshold and the genotype structure is further explored in Fig \ref{fig:2nd_threshold} for the three toy networks across multiple values of $\Delta$. As $\Delta$ increases, immunity becomes wide-reaching in genetic distance, approaching the effects of universal immunity or a universal vaccine. This has the effect of necessitating higher $\beta$ to produce the same $I^*$ as lower values of $\Delta$. Importantly, because of the sum in the denominator of Eq.~\eqref{eq:immune_invasion_threshold}, the immune invasion threshold $\beta_\mathrm{I}$ is not simply set by the diameter of the genotype network (i.e., the maximum value of $x_{ij}$), and is instead set by the entire network structure. While strains maximally distant from each other can much better infect recovered individuals, competition between strains also play an important role: central strains can still infect individuals and grant them better immunity due to their central position in the network. Thus, the network structure plays a nontrivial role in setting the exact value of $\beta_\mathrm{I}$ as determined by Eq.~\eqref{eq:immune_invasion_threshold}.

\begin{figure}[t!]
    \centering
    \includegraphics[width=\linewidth]{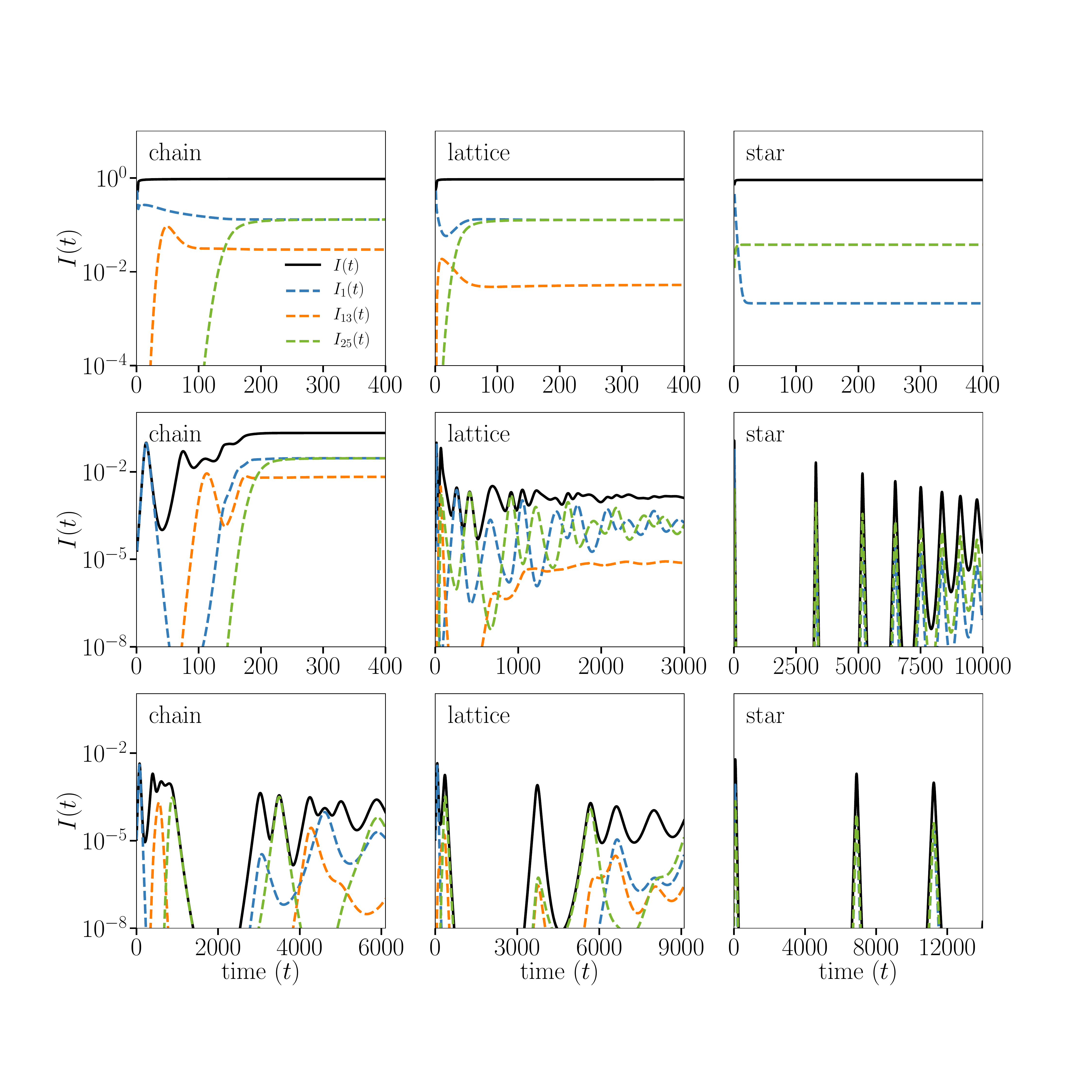}
    \vspace{-1.8cm}
    \caption{Integration of the ODEs on three toy genotype networks --- the chain \textbf{(Left Column)}, square lattice \textbf{(Middle Column)}, and star \textbf{(Right Column)} --- all with 25 strains. We fix the recovery rate $\gamma=1$, the mutation rate $\mu=1/1000$, the waning immunity rate $\alpha =1/5000$ and the transcending immunity $\Delta = 4$ and vary the transmission rate: $\beta = 25$ (\textbf{Top}), $\beta = 1.7$ (\textbf{Middle}), $\beta = 1.1$ (\textbf{Bottom}). The system is initialized with a small fraction $10^{-5}$ of infections on an ``end strain'' (end for the chain, corner for the lattice, leaf for the star). On the chain we see successive activation of all strains, with the system stabilizing once the entire network is explored and evolution reaches a dead-end. The star sees cycles caused by activation of the leaf strains. The lattice is much more interesting, with loops causing a random-like succession of strains to cycle. The dynamics become more interesting for the bottom row, with transmission rates between the epidemic and immune invasion thresholds, with cycles and chaos-like dynamics. The closer we get to the true epidemic thresholds $\beta_c = 1$, the longer the interesting transient dynamics.}
    \label{fig:mid_timeseries}
\end{figure}

\subsection{Rich dynamics in between epidemic thresholds}

Beyond the features of the endemic state, we observe rich prevalence dynamics throughout the epidemic when transmission rates are between the epidemic threshold $\beta_c = 1$ and the immune invasion threshold $\beta_\mathrm{I}\geq \beta_c$. By comparing the top, middle, and bottom rows of Fig \ref{fig:mid_timeseries} we see infection counts throughout the epidemic simulation while the transmission rate lays in different regimes, decreasing from $\beta>\beta_\mathrm{I}$ to values closer to $\beta_c = 1$.

For transmission rate below the immune invasion threshold (bottom two rows), we see oscillations in the overall infection counts across all three networks before converging on an endemic value, resembling a dampened pseudo-chaotic behaviour. Both cyclical and chaotic infection progression have been observed in modelling by Gupta et al., dependent upon the strain structure \cite{Gupta}. The strain structure introduced by the genotype network enables a clear depiction of these phenomena and allows us to show how the network structure itself impacts the range of parameters where chaotic behaviour is expected.

Noting the different time scales shown, the chain rapidly converges on its endemic state while the star undergoes drastic oscillations before convergence. We see variation in infection counts at the strain level, with the infection counts for 3 of the 25 strains shown. At the strain level we see convergence occurring on different time scales within the same network, as well as variability in oscillatory nature. 

In comparison, the top row of Fig \ref{fig:mid_timeseries} shows the rapid convergence on the endemic state when the transmission rate is high ($\beta=25$). There still exists infection localization, as indicated by different endemic infection counts at the strain level, as well as variability in convergence time between strains. However, the oscillatory nature is profoundly absent at transmission rates well above $\beta_\mathrm{I}$. In contrast, as transmission rates are lowered towards $\beta_c = 1$ in the bottom rows of Fig \ref{fig:mid_timeseries}, we see the oscillations preserved but stretched across a broader timescale. Importantly, as the timescale of oscillations is stretched, their minimal values decrease by orders of magnitude. In practice, this shows that any finite size simulations of the dynamics captured by our model would likely lead to strain extinction, with potential to reemerge through mutations. Discrete events are unfortunately not captured in ODE models as they assume continuous values, or infinite population.

\section{Discussion}

The introduction of an underlying genotype network to a multistrain model has demonstrated the emergence of cyclicity, infection localization, and sequential phase transitions, all in one model. Simple mathematical arguments have allowed us to solve for the transitions observed and highlight the nontrivial impact of the structure of the genotype network. Rich infection dynamics are seen between the epidemic threshold and the immune invasion threshold. Altogether, what these results show is that many features of infectious disease dynamics often explained by environmental factors or host behaviour, such as cyclicity \cite{althouse2014epidemic}, unpredictability \cite{scarpino2019predictability} and sequential transitions \cite{allard2017asymmetric}, can also be explained by adding a layer of biological complexity in the form of a genotype network. Our results thus highlight the importance of going beyond the ``one disease, one pathogen" paradigm, with complex dynamics emerging from even the most simple genotype network structures.  

Future work needs to be done to integrate this modelling approach with real genomic data. Likewise, the interplay of our results with the finite size and the contact structure of the host population needs to be investigated; as does the role of strain extinction and emergence. Different modelling approaches will need to be considered, such as explicitly modelling the growth and evolution of the genotype network as it co-evolves (albeit on a different timescale) with the spread of the infectious disease in the host population. Coupling the large modelling literature on growing networks \cite{dorogovtsev2002evolution} with that of network epidemiology \cite{Pastor:2015} should lead to a richer understanding of how networks, both biological and social, impact epidemics. One other temporal feature that should be taken into account is the immune history of individuals. Whereas we currently only consider the most recent infection, the total immune history could grow exponentially with the number of strains considered and therefore represents an important modelling challenge. Finally, this type of model could also be appropriate to reimagine vaccination strategies. The literature on targeted immunization and influential spreaders on networks could then be leveraged \cite{pastor2002immunization,cohen2003efficient,hebert2013global,morone2015influence}, but rather than targeting central individuals the objective would be to best hinder and block the immune evasion of the pathogen as it mutates along its genotype network.

In terms of applying these models to specific scenarios, there is a need for unbiased pathogen genomic data, as well as an understanding of their antigenic properties, to inform models that account for these features using real-world data and to refine the cross-protective immune effects between strains of a pathogen. Similarly, we need more realistic models to take advantage of the growing body of genomic data available and refine the mechanisms driving mutation and immunity. We call for the refinement of immune mechanisms and immune history to allow their incorporation in mathematical disease models. Further understanding of how pathogens explore genotype space, the growth of genotype networks, the role of host immunity towards past strains, and the influence of the above on the fitness landscape of pathogens will better inform models incorporating multiple strains, cross-protective effects, and the evolution of a pathogen.


\end{document}